\documentclass[runningheads]{llncs}
\usepackage{float}
\usepackage{graphicx}
\usepackage{tcolorbox}
\usepackage[linesnumbered, ruled, vlined]{algorithm2e}
\usepackage{amssymb,amsmath}
\usepackage{hyperref}
\hypersetup{
    colorlinks,
    citecolor=blue,
    filecolor=black,
    linkcolor=red,
    urlcolor=green
}
\begin{document}
\title{Utility Driven Job Selection Problem on Road Networks  \thanks{The work of Suman Banerjee is supported by the Seed Grant (SGT-100047) provided by the Indian Institute of Technology Jammu, India. Both the authors have contributed equally in this work.}}
%
%
\author{Mayank Singhal\inst{1} \and
Suman Banerjee \inst{1} }
\authorrunning{Singhal and Banerjee}
%
\institute{Department of Computer Science and Engineering, \\ Indian Institute of Technology Jammu, Jammu \& Kashmir 181221, India. \\
\email{2018ucs0064@iitjammu.ac.in, suman.banerjee@iitjammu.ac.in}\\
}
\maketitle              
\begin{abstract}
In this paper, we study the problem of \textsc{Utility Driven Job Selection} on Road Networks for which the inputs are: a road network with the vertices as the set of Point-Of-Interests (Henceforth mentioned as POI) and the edges are road segments joining the POIs, a set of jobs with their originating POI, starting time, duration, and the utility. A worker can earn the utility associated with the job if (s)he performs this. As the jobs are originating at different POIs, the worker has to move from one POI to the other one to take up the job. Some budget is available for this purpose. Any two jobs can be taken up by the worker only if the finishing time of the first job plus traveling time from the POI of the first job to the second one should be less than or equal to the starting time of the second job. We call this constraint as the temporal constraint. The goal of this problem is to choose a subset of the jobs to maximize the earned utility such that the budget and temporal constraints should not be violated. We present two solution approaches with detailed analysis. First one of them works based on finding the locally optimal job at the end of every job and we call this approach as the \emph{Best First Search Approach}. The other approach is based on the Nearest Neighbor Search on road networks. We perform a set of experiments with real\mbox{-}world trajectory datasets to demonstrate the efficiency and effectiveness of the proposed solution approaches. We observe that the proposed approaches lead to more utility compared to baseline methods. 

\end{abstract}
\section{Introduction} \label{Sec:Intro}
In recent times, the study of road networks has been emerged in the area of data management in general and spatial data management in particular \cite{palanisamy2013road}. Also, we have witnessed a significant development of wireless internet and mobile devices. Hence, for any moving person capturing their location information over time becomes easier. This leads to the availability of trajectory datasets in different repositories \cite{zheng2015trajectory,wang2021survey}. These datasets are effectively used for solving many decision making and recommendation problem that arise in day to day life including point-of-interest recommendation \cite{ying2012urban}, group trip planning query \cite{hashem2013group}, route recommendation \cite{dai2015personalized,qu2019profitable}, travel time prediction \cite{jiang2013travel}, finding most frequent paths \cite{luo2013finding}, driving behavior prediction \cite{liao2020hierarchical} and many more.  
\par Among many one well studied  problem in the context of road networks, is the Food Delivery Problem \cite{yildiz2019provably}. Now-a-days restaurants give the option to the customers for home delivery of the ordered food if the customer is ready to pay the delivery charge. Several startups have grown in this area, namely \emph{Swiggy} \footnote{\url{https://www.swiggy.com/}}, \emph{Zomato} \footnote{\url{https://www.zomato.com/}}, etc. A delivery boy (mentioned as the `worker' in this paper) can earn benefit by serving these jobs. In the literature, there exist some studies related to this problem. Yildiz et al. \cite{yildiz2019provably} studied the problem of meal delivery problem and they proposed a solution methodology  with the assumption of the exact time of order. The objective of their study was to produce a solution that causes maximum number of order delivery. Reyes et al. \cite{reyes2018meal} also studied the meal delivery problem and they developed optimization-based approaches tailored to solve the courier assignment problem and capacity management problem encountered in meal delivery problem. Ji et al. \cite{ji2019alleviating} studied the food delivery problem and solves only the batching problem but not the order carrier assignment problem and the routing problem. Zeng et al. \cite{zeng2019last} studied the problem of last mile delivery problem in a very generic setting but does not consider any time for food preparation. Dai et al. \cite{dai2019o2o} studied the problem of managing the online placed but offline delivered orders by crowed sourced drivers for delivering the foods. 
In this paper, we formulate this problem as a discrete optimization problem and propose two solution approaches. In particular, we make the following contributions in this paper:

\begin{itemize}
\item We study the \textsc{Utility Driven Job Selection} Problem on Road Network where the goal is to select a subset of the jobs originated at different POIs to maximize the utility of the worker.

\item We formulate this problem as a discrete optimization and propose two solution methodologies for this problem with detailed analysis and examples.

\item Both the methodologies have been analyzed in detail to understand their time and space requirement.
\item We conduct an extensive set of experiments with real-life trajectory datasets to show the effectiveness and efficiency of the proposed solution approaches.
\end{itemize}
The rest of the paper is organized as follows. Section \ref{Sec:BPD} describes the background and defines the problem formally. Section \ref{Sec:PSA} contains a detailed description of the proposed solution approaches. Section \ref{Sec:ED} describes the experimental evaluation, and finally, Section \ref{Sec:CFD} concludes our study.  
\section{Background and Problem Definition} \label{Sec:BPD}
In this section, we give the background and define the problem formally. Initially, we start by defining the road network.
\begin{definition}[Road Network]
A road network is defined by a simple, weighted, undirected graph $\mathcal{G}(\mathcal{V}, \mathcal{E}, \omega)$ where the vertex set $\mathcal{V}(\mathcal{G})=\{v_1, v_2, \ldots, v_n\}$ are the set of POIs of a city, and the edge set $\mathcal{E}(\mathcal{G})=\{e_1, e_2, \ldots, e_m\}$ are the road segments joining the POIs and the edge weight function $\omega$ that assigns each edge to its corresponding distance; i.e.; $\omega:\mathcal{E}(\mathcal{G}) \longrightarrow \mathbb{R}^{+}$. 
\end{definition}
We use standard graph\mbox{-}theoretic notations and terminologies from the book by Diestal \cite{diestel2005graph}. As in this study, each vertex of $\mathcal{G}$ represents a POI, hence in the rest of the paper, we use the term `vertex' and `POI' interchangeably.  We denote the number of vertices and edges of $\mathcal{G}$ by $n$ and $m$, respectively. The weight of any edge $(v_iv_j) \in \mathcal{E}(\mathcal{G})$ is denoted by $\omega(v_iv_j)$. If the road network is connected then there must exist at least one path between every pair of vertices, and hence, there must exist a traveling cost as well. For any pair of non-adjacent pairs of POIs, i.e., $(v_iv_j) \notin \mathcal{E}(\mathcal{G})$, we define the travel cost as the minimum cost incurred among all possible paths reaching from $v_i$ to $v_j$. We denote this by  $\mathbb{C}(v_iv_j)$ and defined by the following equation: 

\[
    \mathbb{C}(v_iv_j)= 
\begin{cases}
    \omega(v_iv_j),& \text{if } (v_iv_j) \in \mathcal{E}(\mathcal{G})\\
    \underset{p \in \mathbb{P}(v_iv_j)}{min} \ \underset{(uv) \in \mathcal{E}(p)}{\sum} \ \omega(uv),              & \text{otherwise}
\end{cases}
\]

Here, $\mathbb{P}(v_iv_j)$ is the set of all possible paths reaching from $v_i$ to $v_j$ and for any such path $p$, $\mathcal{E}(p)$ denotes the set of edges that constitute the path $p$. In this study, we consider that for any two POIs $v_i$ and $v_j$, $\mathbb{C}(v_iv_j)=\mathbb{C}(v_jv_i)$. Here, a POI may refer to a hotel, restaurant, cafeteria, and so on. Now-a-days they provide an option of delivering the ordered foods to the residence of the customers and thus creates opportunities for people to earn revenue by serving the food delivery job. We formally define the notion of Job in Definition \ref{Def:2}. 
\begin{definition}[Job] \label{Def:2}
Let, $\mathcal{J}$ be the set of jobs under consideration. Any job $j \in \mathcal{J}$ is defined by a corresponding quintuple $\mathbb{T}_{J}=(j_{id}, j_{p\_id}, j_{u}, t_{j^{s}}, j_{d})$ and the meaning of each symbol is as follows:
\begin{itemize}
\item $j_{id}$: This attribute stores an id corresponding to each job such that each job can be identified uniquely in the job database.
\item $j_{p\_id}$:  This attribute stores the $\texttt{POI\_Id}$ at which this job has been originated.
\item $j_{u}$: This attribute stores the amount of utility to be paid to the worker after finishing this job.
\item $t_{j^{s}}$: This attribute stores the starting time of the job.
\item $j_{d}$: This attribute stores the duration of the job.
\end{itemize}
\end{definition}
It can be observed that in Definition \ref{Def:2} the fifth attribute can also be interpreted as the finishing time (denoted as $t_{j^{f}}$) also as this can be obtained by adding the duration of the job with the starting time. Also, it is natural to consider that for any job $j \in \mathcal{J}$, its finishing time must come after its starting time, and hence, $t_{j^f} > t_{j^s}$. We also have information about the travel time between any two POIs when the worker is traveling using the shortest path. For any two POIs $v_i$ and $v_j$, we denote this quantity as $\mathcal{T}_{v_i,v_j}$. Suppose, $\mathcal{I} \subseteq \mathcal{J}$ be the set of jobs chosen by the worker. Then it is expected that these jobs should satisfy the following two constraints:
\begin{itemize}
\item \textbf{Temporal Constraint:} Consider any two subsequent jobs $j_x$ and $j_{x+1}$ originated at POI $v_x$ and $v_{x+1}$, respectively. So, it is natural that if a worker wants to do the job $j \in \mathcal{J}$ originated at POI $v_j$ and has the starting time $t_{j^{s}}$ then the worker must be  present at the POI $v_j$ on or before $t_{j^{s}}$. So, assume that the worker finished his current assignment at POI $v_x$ and at time $t_x$. Now, if he wishes to choose his next assignment at POI $v_{x+1}$ having the starting time of the job as $t_{x+1}$ then $t_{x+1} \geq t_x+ \mathcal{T}_{x,x+1}$. We call this constraint as the temporal constraint.

\item \textbf{Budget Constraint:} Once the worker finishes his job at POI $v_x$, and moves to the POI $v_{x+1}$ then he needs to pay some price as travel cost. Assume that $j_1$, $j_2$, $\ldots$, $j_k$ is the set of sequenced jobs with their originated POIs at $v_1$, $v_2$, $\ldots$, $v_k$. Let, $\mathbb{C}(v_i,v_j)$ be the cost associated with traveling from $v_i$ to $v_j$. The budget constraint signifies that the total cost for traveling should be less than the allocated budget; i.e.; $\underset{i \in \{1, \ldots, k-1 \}}{\sum} \mathbb{C}(v_i,v_{i+1}) \leq B$. 

\end{itemize}

We also assume that the information about the jobs is available before the worker starts his first job. Now, we state the notion of earned utility for any subset of jobs in Definition \ref{Def:3}. 
\begin{definition}[Earned Utility] \label{Def:3}
For any subset of jobs $\mathcal{I} \subseteq \mathcal{J}$, we define the earned utility by the worker as the total utility that the worker can earn by performing the jobs in Set $\mathcal{I}$. We denote this by $\mathcal{B}(\mathcal{I})$ and define using Equation \ref{Eq:1}.
\begin{equation} \label{Eq:1}
\mathcal{B}(\mathcal{I})= \underset{j \in \mathcal{I}}{\sum} \mathcal{U}(j)
\end{equation} 
Here, $\mathcal{B}(.)$ is the earned utility function that maps each subset of the jobs to the total utility; i.e.; $\mathcal{B}: 2^{\mathcal{J}} \longrightarrow \mathbb{R}_{0}^{+}$, where  $\mathcal{B}(\emptyset)=0$. $\mathcal{U}(j)$ is the utility associated with the Job $j$.
\end{definition}
Suppose, the worker has the budget $B$ to spend for traveling while moving from one job to the next job. Now, the question is which jobs should be chosen by the worker to maximize the earned utility. We call this problem as the \textsc{Utility Driven Job Selection} Problem which is stated in Definition \ref{Def:4}.
\begin{definition}[Utility Driven Job Selection Problem] \label{Def:4} 
Given a road network $\mathcal{G}(\mathcal{V}, \mathcal{E}, \omega)$ and a set of jobs $\mathcal{J}$ originated at different POIs, the goal of this problem is to select a subset of the jobs such that the earned utility is maximized within the allocated budget. 
\end{definition}
From the computational point of view this problem can be posed as follows:

\begin{center}
\begin{tcolorbox}[title=\textsc{Utility Driven Job Selection} Problem, width=12cm] 
\textbf{Input:} Road Network $\mathcal{G}(\mathcal{V}, \mathcal{E}, \omega)$, Set of Jobs $\mathcal{J}$ along with all the required information mentioned in Definition \ref{Def:2}, Budget of the Worker $B$.
\\
\textbf{Problem} Which jobs from the set $\mathcal{J}$ should be chosen by the worker such that the temporal constraint as well as the budget constraint is satisfied and the earned utility is maximized.\\
\textbf{Output:} A subset of the jobs $\mathcal{I} \subseteq \mathcal{J}$.
\end{tcolorbox}
\end{center} 

\section{Proposed Solution Approaches} \label{Sec:PSA}
In this section, we describe the proposed solution approaches for the Utility Driven Job Selection Problem. The first one is the Best First Approach which is stated in the subsequent subsection.
\subsection{Best First Search Approach}  
Assume that the worker $w$ planned to work from time $w^{s}$ till time $w^{f}$. In this approach first the jobs are sorted based on the starting time. The worker takes up the job that comes first in the time horizon after $w^{s}$. If there is a tie then the worker chooses one with the higher utility. From the second job onwards the process goes in the same way with one exception. Now, the worker chooses the `Best' job applying the following criteria. Suppose, the worker is at the POI $v_x$ and this is his $i$-th job that he has finished. Now, let $j_{i+1}$, $j_{i+2}$, $\ldots$, $j_z$ are the set of jobs originated at the POIs $v_{i+1}$, $v_{i+2}$, $\ldots$, $v_{z}$ respectively satisfies the temporal constraint. Now, the worker will try to choose  the job $j \in \{j_{i+1}, j_{i+2}, \ldots, j_{z}\}$ originated at the POI $v_y$ if $\mathcal{U}_{J} > \mathbb{C}(v_x,v_y)$. If there are more than one jobs that satisfies this criteria then the worker will choose one with the highest $(\mathcal{U}(j)- \mathbb{C}(v_x,v_j))$ value stated in 
\begin{equation}
j^{*}_{i+1}= \underset{j \in \{j_{i}, j_{i+1}, \ldots, j_{y}\}}{argmax} \ \mathcal{U} (j)- \mathbb{C}(v_x,v_j)
\end{equation}
The worker will stop when either the budget of the worker is exhausted or 
the completion time of the worker is reached or both. This idea is described in Algorithm \ref{Algo:1}. For the simplicity, we also assume that the starting point of the worker is also a POI in the road network $\mathcal{G}$.
\paragraph{Description of Algorithm \ref{Algo:1}} We assume that the number of jobs in the database is $k$; i.e.; $|\mathcal{J}|=k$. $\mathcal{J}^{w}$ contains the subset of the jobs chosen for the worker and initially assigned to empty set. $\mathcal{U}^{w}$ denotes the utility earned by the worker $w$ which is initialized to $0$. $v_c$ is the starting POI of the worker. As the worker starting from time  $w^{s}$ hence those jobs whose starting time is before $w^{s}$ has no role to play. Hence, we delete all such jobs. Now, Line $6$ to $15$ Algorithm \ref{Algo:1} is executed for selecting the first job. Even before that, we sort all the jobs based on the starting time. Until any one of the termination criteria is met we keep on scanning the jobs, we pick up the first job if it satisfies the temporal constraints. If a job does not obey this constraint then we delete this job as it can be done by the worker. As soon as one job is found, we exit from the \texttt{while} loop of Line $6$. While exiting, we set the current time as the finishing time of the job selected as the first job. Also, the earned utility by the worker, the set of selected jobs, and the budget are updated accordingly. Subsequent jobs have been chosen and the selection process goes like this. Until one of the exiting criteria is met the remaining jobs are processed. Without loss of generality, we assume that the job in the $i$-th index  (denoted as $j_{i}$) of the sorted list $\mathcal{J}$ and its originating time and finishing time are $t_{j^{s}_{i}}$. Now, we notice the jobs that have the same originating time as the $(i+1)$-th job and we keep in the list $\mathcal{J}^{z}$. From these jobs we choose one that maximizes the difference between the utility earned by performing the job and the traveling cost inured is maximized. Once a job is selected accordingly all the parameters are updated. Finally, the set of selected jobs and the earned utility are returned. 
\begin{algorithm}[!htb]
  \DontPrintSemicolon
  \KwIn{A Road Network $\mathcal{G}(\mathcal{V}, \mathcal{E}, \omega)$, The Job Database $\mathcal{J}$, The Worker $w$ with Its Initial Location, starting and ending time, Budget $B$.}
  \KwOut{A Subset of the Jobs $\mathcal{J}^{w} \subseteq \mathcal{J}$.}
  $[w^{s}, w^{f}] \longleftarrow \text{Working interval of the worker}$\;
  $v_p \longleftarrow \text{Starting location of the worker}$\;
  $\mathcal{J}^{w} \longleftarrow \emptyset$; $\mathcal{U}^{w} \longleftarrow 0$; $t \longleftarrow w^{s}$; $i \longleftarrow 1$\;
 $\mathcal{J} \longleftarrow \mathcal{J} \setminus \{j: t_{j^{s}} < w^{s} \}$\;
 $ \mathcal{J} \longleftarrow \text{Sort the jobs based on the } t_s \text{ value}$\;
  \While{$B > 0 \text{ and } t \leq t_{j^{s}_{k}}$}{
  \For{$a=1 \text{ to }k$}{
  \eIf{$t+\mathcal{T}(v_x,v_a) \leq t_{j^{s}_{a}} \text{ and } \mathbb{C}(v_p,v_a) \leq B$}{
  $\mathcal{J}^{w} \longleftarrow \mathcal{J}^{w} \cup \{j_{a}\}$; $\mathcal{U}^{w} \longleftarrow \mathcal{U}^{w} + \mathcal{U}(j_a)$\;
  
  $t \longleftarrow t_{j^{f}_{a}}$; $B \longleftarrow B - \mathbb{C}(v_p,v_a)$\;
  
  $\texttt{break}$\;
  }
  {
  $\mathcal{J} \longleftarrow \mathcal{J} \setminus \{j_{a}\}$
  }
  }
  \If{$|\mathcal{J}^{w}|==1$}{
  $\texttt{break}$\;
  }
  }
  \While{$B > 0 \text{ and } t \leq t_{j^{s}_{k}}$}{
  $t^{s}_{j_{a+1}} \longleftarrow \text{Starting time of the job }j_{a+1}$\;
  $\mathcal{J}^{z} \longleftarrow j_{a+1}$\;
  \For{$z=a+2 \text{ to }k$}{
  \eIf{$t^{s}_{j_{a+1}}==t^{s}_{z}$}{
  $\mathcal{J}^{z} \longleftarrow \mathcal{J}^{z} \cup \{j_{z}\}$\;
  }
  {
  $\texttt{break}$\;
  } 
  }
  $j^{*} \longleftarrow \underset{j \in \mathcal{J}^{z} \text{ and } t+ \mathcal{T}(v_a,v_z) \leq t^{s}_{j}}{argmax} \ \mathcal{U}(j) - \mathbb{C}(j_{a}, j_{z})$\;
  $\mathcal{J}^{w} \longleftarrow \mathcal{J}^{w} \cup \{j^{*}\}$; $B \longleftarrow B - \mathbb{C}(j_{a}, j_{z})$\;
  
  $t \longleftarrow j^{f}_{z}$; $\mathcal{U}^{w} \longleftarrow \mathcal{U}^{w} + \mathcal{U}(j^{*})$; $a \longleftarrow z$\;
  }
  $\texttt{return } \mathcal{J}^{w}, \mathcal{U}^{w} $\;
  
  \caption{\textsc{Best First Approach} for the \textsc{Utility Driven Point-Of-Interest Selection} Problem}
  \label{Algo:1}
\end{algorithm}
\paragraph{Analysis of Algorithm \ref{Algo:1}} Now, we analyze Algorithm \ref{Algo:1} to understand its time and space requirement. The statements mentioned from Line $1$ to $3$ are merely input statement and requires $\mathcal{O}(1)$ time. Line $4$ can be implemented by just a single scan of the jobs and it takes $\mathcal{O}(k)$ time. Sorting the jobs require $\mathcal{O}(k \log k)$ time. Now, the running time of the remaining portion of the algorithm will depend on how many times the \texttt{while} loops in Line $6$ and $15$. Assume that $\mathbb{C}_{min}$ denotes the minimum traveling cost among all possible POI pairs; i.e.; $\mathbb{C}_{min}=\underset{v_x,v_y \in \mathcal{V}(\mathcal{G})}{min} \mathbb{C}(v_x,v_y)$. If the budget of the worker is $B$, then the maximum number of jobs that cannot be done by the worker $\frac{B}{\mathbb{C}_{min}}$. Consider the sorted list of jobs obtained after the execution of Line $5$. Let, $t_{j^{s}_{1}}$ and $t_{j^{s}_{k}}$ denotes the originating time of the first and last job and this difference between $t_{j^{s}_{k}}$ and $t_{j^{s}_{1}}$ is denoted by $\Delta$; i.e.; $\Delta=t_{j^{s}_{k}}-t_{j^{s}_{1}}$. Now, it is easy to observe that the both the \texttt{while} loops can run at most $min(\frac{B}{\mathbb{C}_{min}}, \Delta)$ times. We denote this quantity as $r$. Now, it is easy to observe that the \texttt{for} loop in Line $7$ can runs at most $\mathcal{O}(k)$ times. Also, we can observe that the condition checking statements at Line $8$ and other statements from Line $9$ to $15$ requires $\mathcal{O}(1)$ time. So, the running time from Line $6$ to $15$ requires $\mathcal{O}(r \cdot k)$ time. It is easy to observe that the statements of Line $16$ and $17$ requires $\mathcal{O}(1)$. Let, for any $t \in [t_{j^{s}_{1}}, t_{j^{s}_{k}}]$, $\mathcal{J}_{t}$ denotes the set of jobs with $t$ as the  Let, $\mathcal{J}^{max}_{t}$ denotes the maximum number of jobs at any particular time in the time interval $[t_{j^{s}_{1}}, t_{j^{s}_{k}}]$ as originating time; i.e.; $\mathcal{J}^{max}=\underset{t \in [t_{j^{s}_{1}}, t_{j^{s}_{k}}]}{argmax} \ \mathcal{J}_{t}$.  It is easy to observe that the \texttt{for} loop in Line $18$ can run at most $\mathcal{O}(\mathcal{J}^{max})$ many times. The condition checking statement of the \texttt{if} statement at Line $19$ and statements in Line $20$ and $22$ will take $\mathcal{O}(1)$ time to execute. We can easily observe that the instruction in Line $23$ can be implemented in $\mathcal{O}(\mathcal{J}_{t})$ time. All the remaining instructions in Line $24$ and $25$ will take $\mathcal{O}(1)$ time. Hence, the time requirement from Line $15$ to $25$ requires $\mathcal{O}(r \cdot \mathcal{J}_{t})$ time. Hence, the total time requirement  of Algorithm \ref{Algo:1} requires $\mathcal{O}(k + k \log k +r \cdot k + r \cdot \mathcal{J}_{t}) = \mathcal{O}(k \log k +r \cdot k + r \cdot \mathcal{J}_{t})$. Other than the inputs extra space taken by Algorithm \ref{Algo:1} is to store the lists $\mathcal{J}^{w}$ and $\mathcal{J}^{z}$ which requires $\mathcal{O}(k)$ space. All the remaining variables used in Algorithm \ref{Algo:1} requires $\mathcal{O}(1)$ space. Hence, the space requirement of Algorithm \ref{Algo:1} is of $\mathcal{O}(k)$. Note that in Algorithm \ref{Algo:1} we consider that pairwise travel cost and time between is part of input itself.  The final statement is mentioned in Theorem \ref{Ref:Th:1}.
\begin{theorem} \label{Ref:Th:1}
The running time and space requirement of Algorithm \ref{Algo:1} is of $\mathcal{O}(k \log k +r \cdot k + r \cdot \mathcal{J}_{t})$ and $\mathcal{O}(k)$, respectively.
\end{theorem}
It is important to note that we are not considering the time requirement for computing the shortest path distance matrix as this can be computed offline.
\subsection{Solution Approach based on Nearest Neighbor Search}
This method relies on computation of the nearest neighbor on road network. In this method, the nearest neighbor is searched based on the priority of the job which is stated in Definition \ref{Ref:Priority}. The priority function has been designed by observing the following facts:
\begin{itemize}
\item If the sum of traveling time for moving from current POI to the POI where the job has been originated and time requirement for performing the job is more then the priority of the job should be less.
\item If the utility obtained by performing the job is higher then the priority of the job will also be more. 
\end{itemize}
 The priority can be recognized by a function $\mathbb{P}$ that maps every tuple of job and POI to its priority value which is a positive real number, i.e., $\mathbb{P}: \mathcal{J} \times \mathcal{V}(\mathcal{G}) \longrightarrow \mathbb{R}^{+}$.    
\begin{definition}[Priority of the Job] \label{Ref:Priority}
For any job $j \in \mathcal{J}$ when the worker is at the POI $v_i$ we denote its priority by $\mathbb{P}_{i}(j)$ and defined by Equation \ref{Eq:Priority}.
\begin{equation} \label{Eq:Priority}
\mathbb{P}_{i}(j)= \frac{\mathcal{U}(j)-\mathbb{C}(v_i,v_j)}{\mathcal{T}(v_{i},v_{j})+(t_{j^{f}}-t_{j^{s}})}
\end{equation} 
\end{definition}
From Definition \ref{Ref:Priority} the following two points can be observed.
\begin{itemize}
\item From Equation \ref{Eq:Priority}, it can be noticed that the priority function depends on the current POI of the worker.
\item Secondly, for any particular POI the highest priority job will be one which has the higher utility and lower time requirement for completion of the job. 
\end{itemize}
\paragraph{Description of Algorithm \ref{Algo:2}} Now, we state the working principle of Algorithm \ref{Algo:2}. First four lines of Algorithm \ref{Algo:2}  is same as Algorithm \ref{Algo:1}. Subsequently, we call the current location of the worker as previous because possibly the worker will move subsequently it will become the previous POI for him. Now, the job selection process starts. Until the budget of the worker is exhausted and the current time is less than the maximum working time of the worker the following steps are repeated. First, we take out the POIs where the Jobs in the database has been originated. Based on the current location of the worker we compute the priority of the jobs using Equation \ref{Eq:Priority}. Next, the nearest neighbor is computed from the current  location of the worker and the locations of the current jobs in $\mathcal{J}$. Here, we point out that the nearest neighbor is computed based on the previously computed priority values. Now, the job originated at the POI corresponding to the nearest POI based on the priority value. Next thing is to be checked whether this job satisfies the temporal constraint or not. If this constraint is satisfied then this job is taken up by the worker and the utility associated with this job is earned by the worker. The budget of the worker is reduced by the amount of traveling cost from the previous to next POI. The current time is updated as the finishing time of the job and \emph{previous} POI has been updated by the next. If the job does not satisfy the temporal constraint then the job is removed from the job database. After exiting from the \texttt{while} loop we get the set of jobs along with the earned utility.

\begin{algorithm}[!htb]
  \DontPrintSemicolon
  \KwIn{A Road Network $\mathcal{G}(\mathcal{V}, \mathcal{E}, \omega)$, The Job Database $\mathcal{J}$, The Worker $w$ with Its Initial Location, starting and ending time, Budget $B$.}
  \KwOut{A Subset of the Jobs $\mathcal{J}^{w} \subseteq \mathcal{J}$.}
  $[w^{s}, w^{f}] \longleftarrow \text{Working interval of the worker}$\;
  $v_p \longleftarrow \text{Starting location of the worker}$\;
  $\mathcal{J}^{w} \longleftarrow \emptyset$; $\mathcal{U}^{w} \longleftarrow 0$; $t \longleftarrow w^{s}$;\;
 $\mathcal{J} \longleftarrow \mathcal{J} \setminus \{j: t_{j^{s}} < w^{s} \}$\;
 $ \mathcal{V}(\mathcal{G}) \longleftarrow \mathcal{V}(\mathcal{G}) \setminus \{v_p\}$\;
 $previous \longleftarrow v_p$\;
  \While{$B > 0 \text{ and } t \leq t_{j^{s}_{k}}$}{
  $V \longleftarrow \text{POIs of the Jobs in } \mathcal{J}$\;
  $\mathcal{P} \longleftarrow \text{ Calculate the Priority using Equation } \ref{Eq:Priority}$\;
  $next \longleftarrow \text{Nearest Neighbor Search }(v_p, V) \text{ based on the Priority Value}$\;
  \eIf{$\mathcal{T}(previous,next) + (t_{j^{s}_{x}}- t_{j^{s}_{x}}) \leq w^{f}$}{
  $\mathcal{J}^{w} \longleftarrow \mathcal{J}^{w} \cup \{j_x\}$\;
  $\mathcal{U}^{w} \longleftarrow \mathcal{U}^{w} + \mathcal{U}(v_j)$\;
 
  $B \longleftarrow B- \mathbb{C}(previous,v_x)$\;
  $t \longleftarrow t_{j^{f}_{x}}$; $previous \longleftarrow \ next$\;
  }
  {
  $\mathcal{J} \longleftarrow \mathcal{J} \setminus \{j_x\}$\;
  }
  }

  $\texttt{return } \mathcal{J}^{w}, \mathcal{U}^{w} $\;
  
  \caption{\textsc{Nearest Neighbor Search Approach} for the \textsc{Utility Driven Point-Of-Interest Selection} Problem}
  \label{Algo:2}
\end{algorithm}

\paragraph{Analysis of Algorithm \ref{Algo:2}} Now, we proceed to analyze Algorithm \ref{Algo:2}. First three lines of Algorithm \ref{Algo:2} is merely the inputs and as mentioned previously removing the jobs having the originating time before the starting time of the worker requires $\mathcal{O}(k)$ time. Line $6$ requires $\mathcal{O}(1)$ time to execute. Like Algorithm \ref{Algo:1}, the running time of Algorithm \ref{Algo:2} will also depends upon how many  times the \texttt{while} loop at Line $7$ is executing. As mentioned in Algorithm \ref{Algo:1}, the number of times the \texttt{while} loop is running is of $\mathcal{O}(r)$ where $r=min(\frac{B}{\mathbb{C}_{min}}, \Delta)$. Now, POIs corresponding to the jobs can be obtained by a single scan which requires $\mathcal{O}(k)$ time. From Equation \ref{Eq:Priority}, it can be easily observed that for any particular job the computation of its priority requires $\mathcal{O}(1)$ time. Hence, for all the jobs computing the priority requires  $\mathcal{O}(k)$ time. Subsequently, choosing the nearest neighbor will take $\mathcal{O}(k \log k)$ in the worst case. All the remaining statements of Algorithm \ref{Algo:2} will take $\mathcal{O}(1)$ time. So, the total time requirement is of $\mathcal{O}(k \log k+ r(k+))$ It is easy to observe that the space requirement of Algorithm \ref{Algo:2} is of $\mathcal{O}(k)$ only. Hence, Theorem \ref{Th:2} holds.

\begin{theorem} \label{Th:2}
The running time and space requirement of Algorithm \ref{Algo:2} is of $\mathcal{O}(k \log k)$ and $\mathcal{O}(k)$, respectively.
\end{theorem}   

\section{Experimental Details} \label{Sec:ED}
In this section, we describe the experimental evaluations of the proposed solution approaches. Initially, we start by describing the datasets.
\paragraph{\textbf{Description of the Datasets}} 
In this study, we use three trajectory datasets, namely Europe Road Network (ERD), Minnesota Road Network (MNR), and Oldenburg Road Network (OBR) \cite{nr}. The first two have been downloaded from \url{https://networkrepository.com/road.php} and the third one has been downloaded from \url{https://www.cs.utah.edu/~lifeifei/SpatialDataset.htm}. Table \ref{tab:Dataset} contains the basic statistics of the datasets.

\begin{table}
\centering
  \caption{Basic Statistics of the Datasets}
  \label{tab:Dataset}
  \begin{tabular}{ccccc}
    \hline
    Dataset Name & $n$ & $m$ & Density & Avg. Degree \\
    \hline
    \textbf{Europe Road Network} & 1.2K & 1.4K & 0.00205794 & 2 \\
    \textbf{Minnesota Road Network} & 2.6K & 3.3K & 0.000946754 & 2 \\
   \textbf{City of Oldenburg} & 6.1K & 7K & 0.000377 & 2.302 \\
  \hline
\end{tabular}
\end{table}

\paragraph{\textbf{Algorithms in our Experiments}}
In this paper, we compare the performance of the proposed solution methodologies with the following two baseline methods:
\begin{itemize}
\item \textbf{Random (RANDOM)}: In this method, the worker randomly picks up a job and checks its feasibility (both the budget constraint and temporal constraint). If they are satisfied then the job is taken up by the worker else the worker searches for another job  until a suitable one is found.
\item \textbf{Utility-Based Greedy (U-GREEDY)}: In this method, the worker chooses the job that satisfies both the constraints and has the highest utility value. In this method, the objective is to become greedy with respect to the utility of the jobs.
\end{itemize}
Other than this we have the proposed solution methodologies: Best First Search(BFS) Approach, and Nearest Neighbor(NN) Approach. Additionally, For the second method, we create a variant of it where we consider that the underlying spatial database is indexed with $\mathcal{R}$-Tree. Due to space limitations, we can't discuss more on it. However, interested readers may look into \cite{brakatsoulas2002revisiting}. 
\paragraph{\textbf{Experimental Setup}}
We adopt the following experimental setup. In all three datasets, between every pair of vertices distance value is available. We assume that the travel time is proportional to the distance. We generate the travel time by multiplying $0.2$ with the distance. We consider the time interval as $[1, 5000]$ and we create $200$, $400$, and $800$ jobs. For each job, we assign its utility from the interval $[9K, 12K]$. In our experiments, our goal is to check if we increase the number of jobs then how the number of performed jobs and the earned benefit is changing.

\begin{figure*}[h!]
\centering
\begin{tabular}{cc}
\includegraphics[scale=0.2]{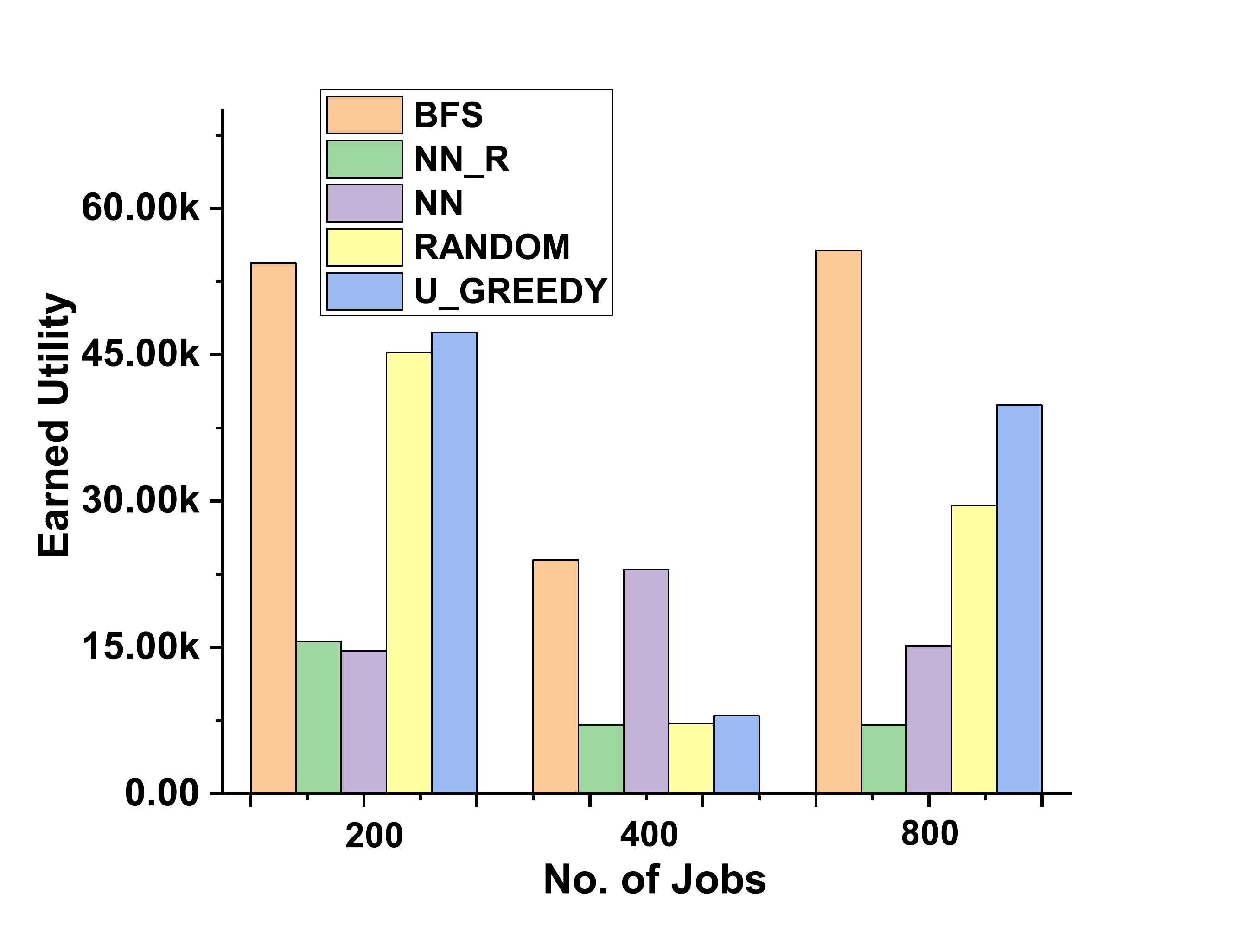} & \includegraphics[scale=0.2]{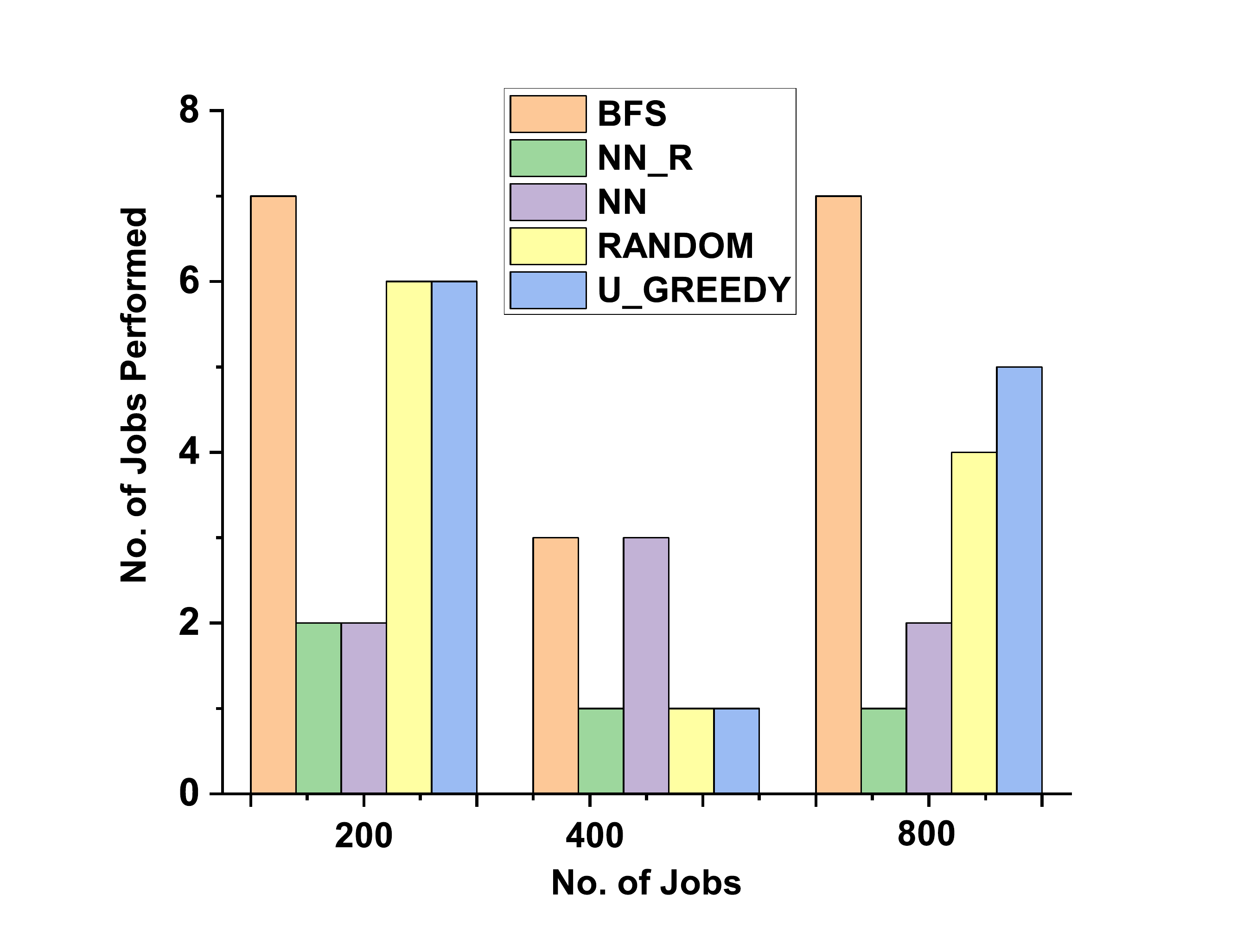} \\
(a) ERD(\# Jobs Vs. Earned Utility)  & (b) ERD(\# Jobs Vs. Performed Jobs) \\
\includegraphics[scale=0.2]{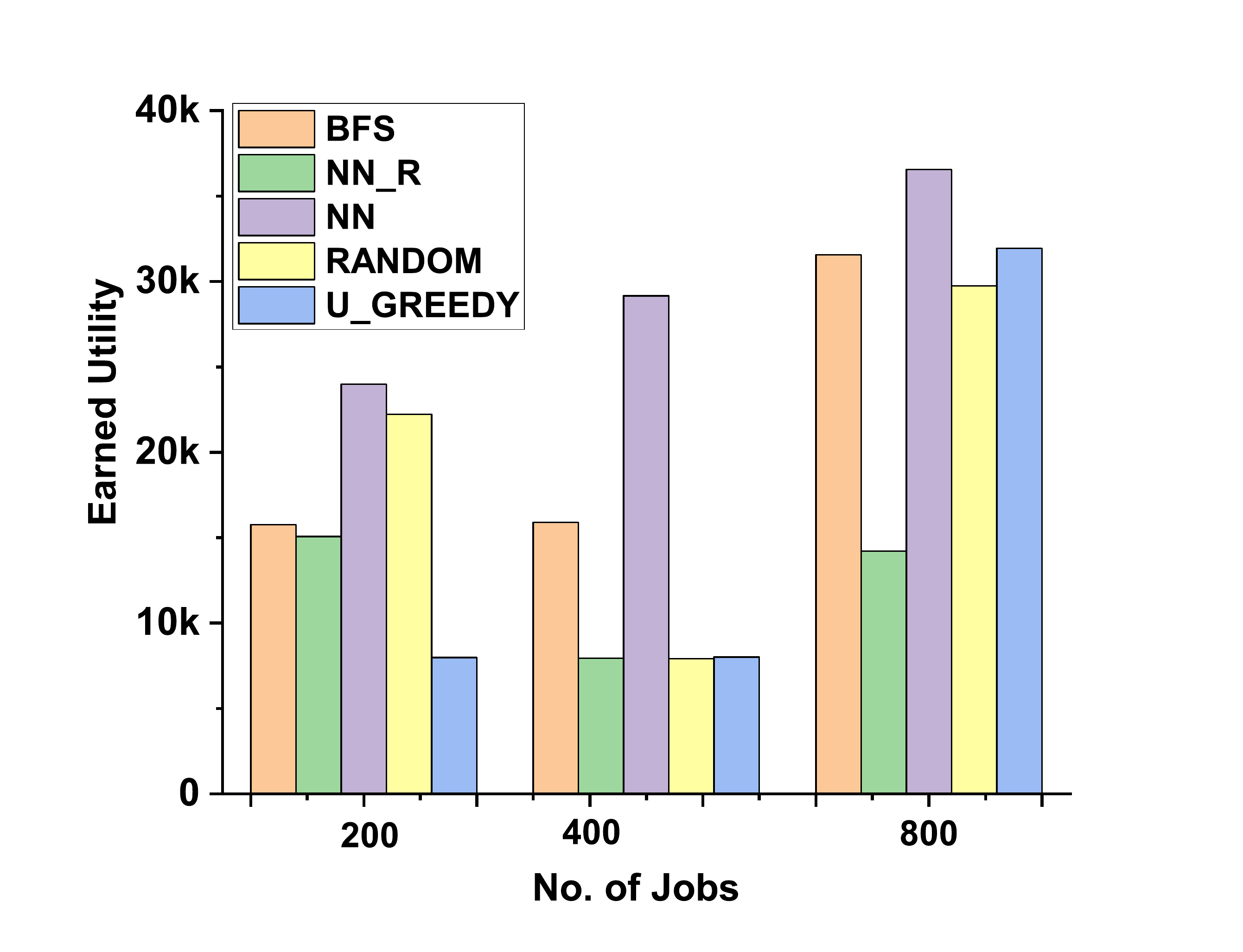} & \includegraphics[scale=0.2]{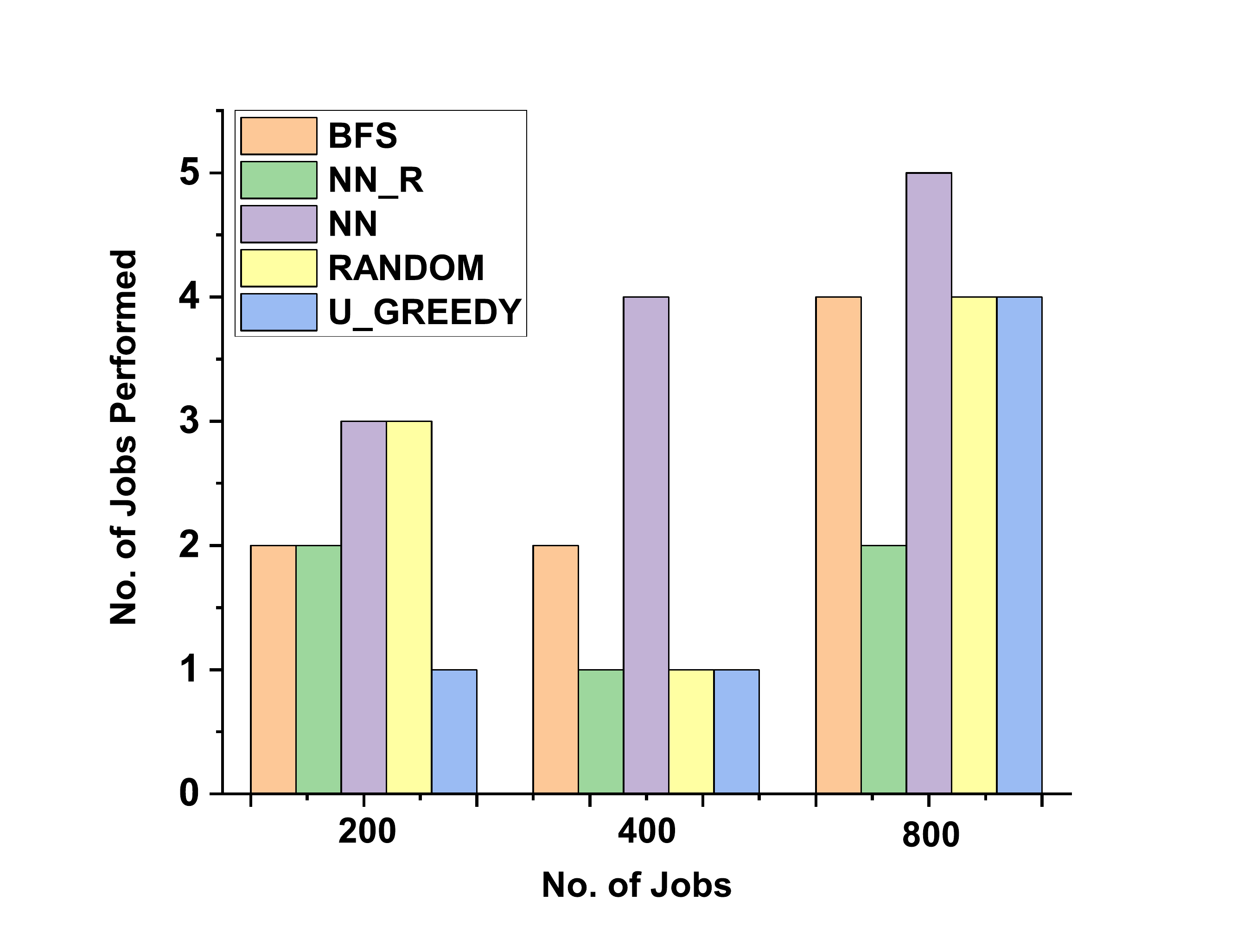} \\

(c) MNR(\# Jobs Vs. Earned Utility) & (d) MNR(\# Jobs Vs. Performed Jobs)  \\
\includegraphics[scale=0.2]{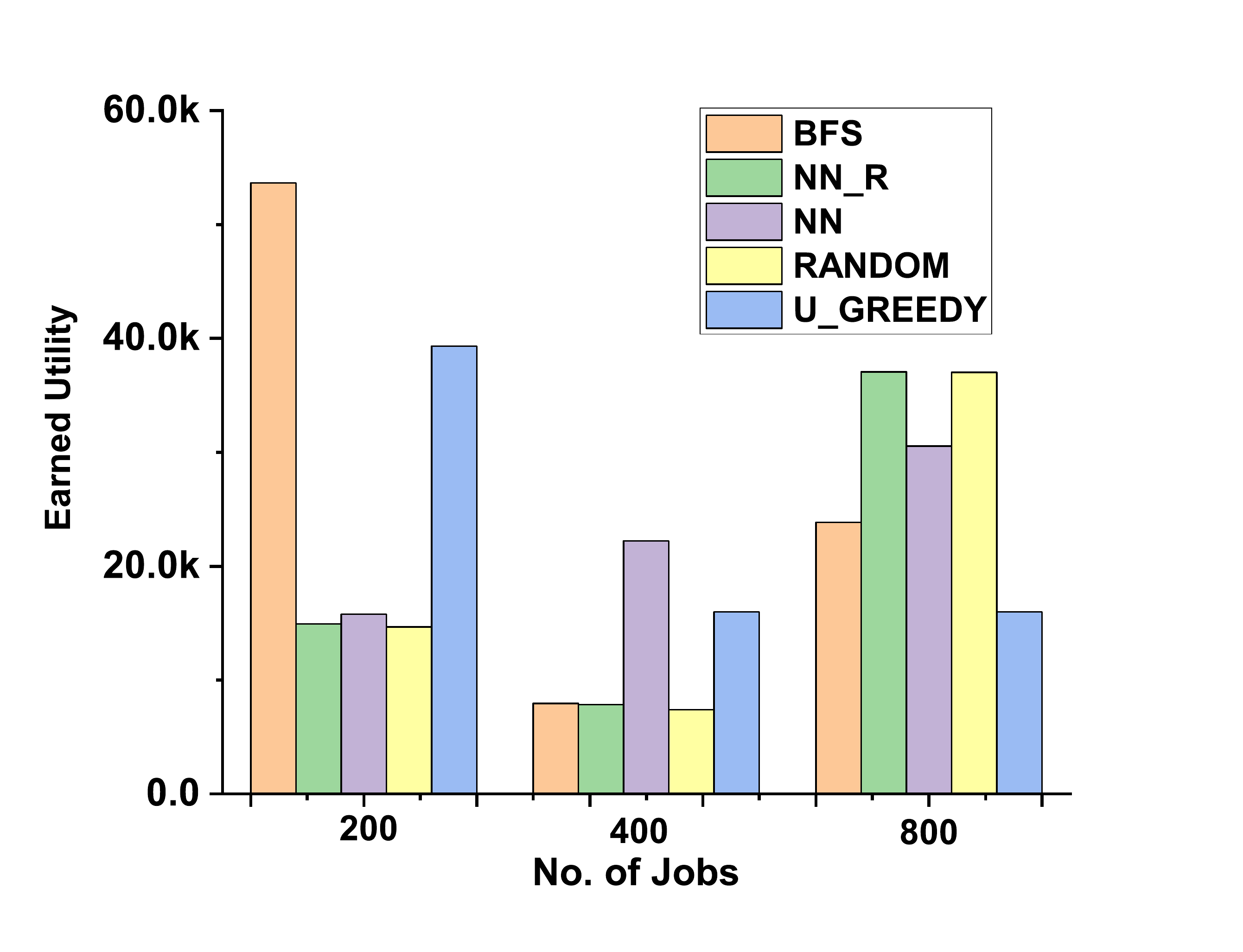} & \includegraphics[scale=0.2]{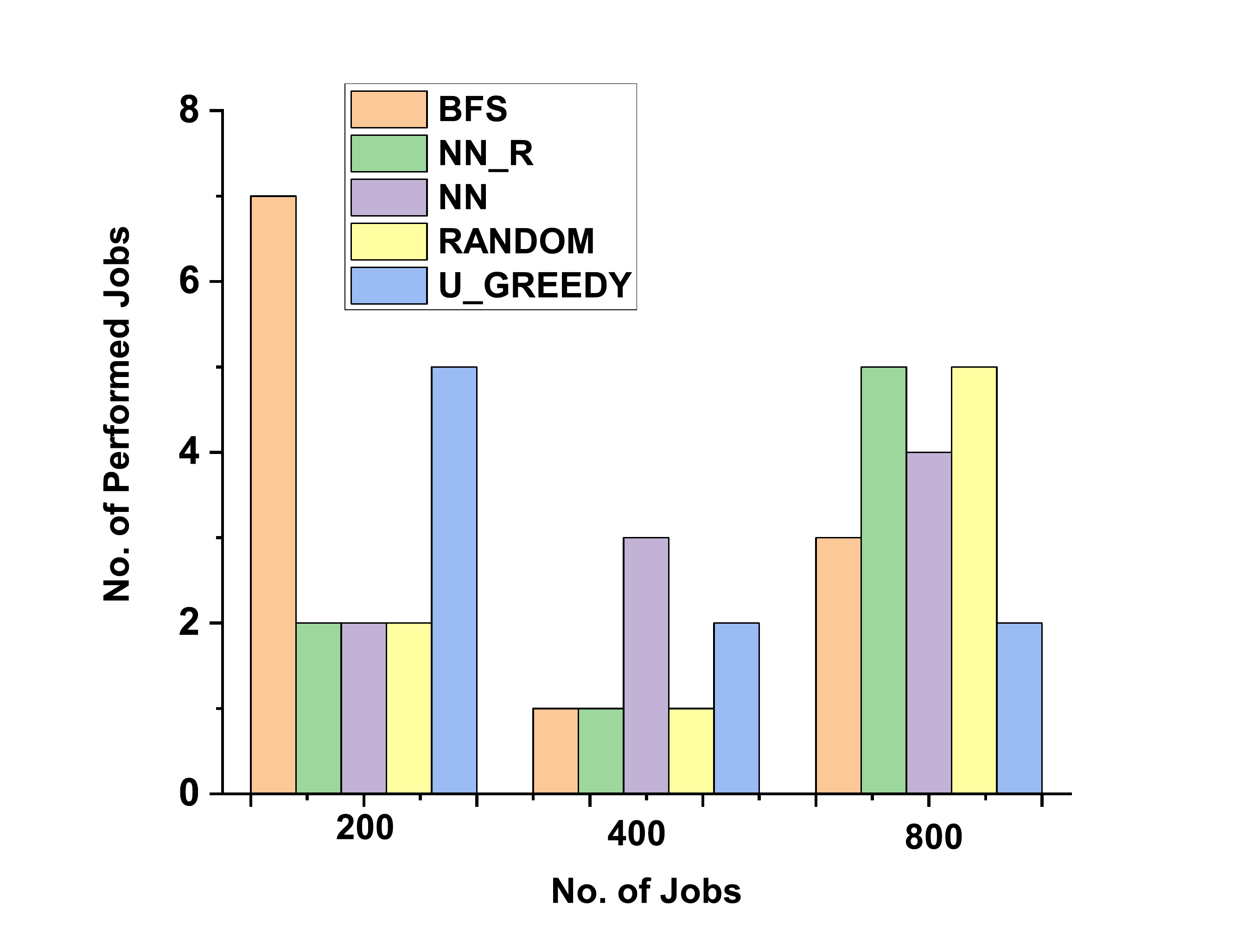} \\
(e) OBR(\# Jobs Vs. Earned Utility) & (f) OBR(\# Jobs Vs. Performed Jobs) \\
\end{tabular}
\caption{No. of Jobs Vs. Earned Utility and No. of Jobs Vs. No. of Performed Jobs plots for three datasets}
\label{Fig:Results}
\end{figure*}

\paragraph{\textbf{Experimental Results with Discussions}}
In this section, we describe the experimental results with detailed discussions. Figure \ref{Fig:Results} shows the plots of No. of Jobs Vs. Earned Utility and No. of Jobs Vs. No. of Performed Jobs for all the datasets. From the figure, we can observe that for all the datasets in most of the instances the proposed solution approaches lead to the more amount of earned utility compared to the baseline methods. Now, we explain the obtained results dataset wise. For the Europe Road network Dataset, we observe that among the proposed methodologies the jobs selected by the Best First Approach leads to more amount of earned utility. As an example when the number of jobs are $200$, among the proposed methodologies the Best First Approach leads to the earned utility of $54362$. However, among the baseline methods, the Utility-based Greedy approach leads to the earned utility of $47319$, which is approximately $15 \%$ more compared to the baseline method. One important observation in this dataset is that the earned utility is not proportional to the number of available jobs. As an example, for the Best First Search method when the number of jobs has been increased from $200$ to $400$, the earned utility is decreased from $54362$ to $23935$. The reason behind this is that increasing the number of jobs does not ensure that it will increase the number of performed jobs and that is what has been reflected in our experiments (Fig. \ref{Fig:Results} (b), (d), (f)).

\par For the Minnesota Road Network dataset also the observations are quite similar. When the number of jobs are $200$, the earned utility by the Nearest Neighbor Approach is $23986$. However, among the baseline methods,  the performance of the Random method is better than the Utility-based Greedy and the earned utility by the Random Method is $22219$. This is approximately $8 \%$ more. In this dataset also, we observe when the number of jobs are increased from $200$ to $400$, the earned utility values are decreased for the Nearest Neighbor and Utility-based Greedy methods. For the Nearest Neighbor Method, the decrement is from $15059$ to $7940$ and the same for the Random Method drop down from $22219$ to $7915$. For the City of Oldenburg Road Network dataset also the observations are quite similar. Among the proposed methodologies, when the number of jobs are $200$, the Best First Search Method leads to the maximum amount of earned utility and the amount is $53663$ and the same for the Utility-based Greedy Method is $39330$. This is approximately $35 \%$ more. Finally, from our  experiments, we list out the following two key observations: 
\begin{itemize}
\item The proposed solution approaches lead to more amount of earned benefit compared to baseline methods.
\item Increase in number of jobs does not ensure the increase in the earned benefit.
\end{itemize}

\section{Conclusion and Future Research Directions} \label{Sec:CFD}
In this paper, we have studied the problem of Utility Driven Job Selection where the goal is to choose a subset of the jobs originated at different POIs of a road network such that the utility earned by performing the jobs is maximized and also the total cost for in-between travel cost by the worker is bounded by the given budget. For this problem, we develop two solution approaches namely the Best First Search approach, and the Nearest Neighbor Search Approach. We have analyzed both the methods to understand their time and space requirements. Subsequently, we conduct several experiments to evaluate the proposed methodologies. Our future work on this problem will concentrate on developing more scalable algorithms along with more realistic constraints such as the number of people required for completion of a job is more than one, the existence of more than one worker in the system and so on.  

 \bibliographystyle{splncs04}
 \bibliography{Paper}
%
%
%
%
%
\end{document}